\documentclass[twocolumn,showpacs,prb,amsmath]{revtex4}
\usepackage{epsfig,graphicx,bm}

\def\be{\begin{equation}}
\def\ee{\end{equation}}
\def\e#1{\label{#1}\end{equation}}
\def\bea{\begin{eqnarray}}
\def\eea{\end{eqnarray}}
\def\ea#1{\label{#1}\end{eqnarray}}
\def\bes#1{\begin{subequations}\label{#1}}
\def\ese{\end{subequations}}
\def\t{{\bf t}}
\def\r{{\bf r}}

\begin{document}
\title{Flying microwave qubits with nearly perfect transfer efficiency}
\author{Alexander N.\ Korotkov}
 \affiliation{Department of
Electrical Engineering, University of California, Riverside,
California 92521}
\date{\today}

\begin{abstract}
We propose a procedure for transferring the state a microwave qubit
via a transmission line from one resonator to another resonator,
with a theoretical efficiency arbitrarily close to 100\%. The
emission and capture of the microwave energy is performed using
tunable couplers, whose transmission coefficients vary in time.
Using the superconducting phase qubit technology and tunable
couplers with maximum coupling of 100 MHz, the procedure with
theoretical efficiency $\eta=0.999$ requires a duration of about 400
ns (excluding propagation time) and an ON/OFF ratio of 45. The
procedure may also be used for a quantum state transfer with optical
photons.
    \end{abstract}


  \maketitle

\section{Introduction}

   Rapid progress in experiments with superconducting qubits \cite{sc-qubits}
confirms their importance for quantum information processing.
However, all such experiments so far are confined to one chip inside
a dilution refrigerator, without a possibility to transfer quantum
information from chip to chip or over a longer distance between the
refrigerators. In particular, the superconducting experiment on the
Bell inequality violation \cite{Bell} has been done with the
locality loophole. A natural way of passing quantum information from
site to site is by using ``flying qubits'' represented by single
photons (or more precisely, superpositions of one and zero photons).
For a superconducting qubit we may think about emission of a
microwave photon, which propagates through a lossless
superconducting waveguide and is then ``captured'' by another qubit.
The technology of superconducting qubits coupled to microwave
resonators based on coplanar waveguides is now well
developed.\cite{coplanar} Since the resonators have better coherence
time than the qubits, it may be beneficial to transfer quantum
information between two resonators instead of coupling qubits to the
transmission line directly.

    The transfer of quantum information from site to site may seem
much easier in optics; however, this is not the case. Even though
optical photons easily propagate in fibers, it is not easy to
capture a photon for further information processing, without
destroying it. An important idea is to use trapping of photon states
in atomic ensembles. \cite{Lukin} In this way entanglement between
remote atomic ensembles can be established and then used to transfer
quantum information; however, general processing of the quantum
information by linear optics means \cite{KLM} is problematic because
of its indeterministic nature. Some other approaches (e.g., Refs.\
\onlinecite{other}) may also be useful for the transfer of quantum
information over large distances, but in any case such transfer is
not simple.

    A promising idea for a quantum state transfer between two
identical oscillators of either optical or microwave range of
frequency via a transmission line was put forward by Jahne, Yurke,
and Gavish. \cite{Yurke} In their scheme the coupling between the
emitting oscillator and the transmission line changes in time. It
was shown \cite{Yurke} that with a specific time dependence of the
coupling, the fidelity $F$ of the transfer can be made arbitrarily
close to 100\%. Unfortunately, the required ON/OFF ratio for the
coupler is quite large, ON/OFF$\gg 1/(1-F)$, and the duration of the
procedure $T$ (excluding the propagation time) is quite long, $T\gg
Q_{\rm min}^{\rm em} /[\omega (1-F)] \ln (1-F)^{-1}$, where $\omega$
is frequency of the oscillators and $Q_{\rm min}^{\rm em}\gg 1$ is
the minimum ``loaded'' quality factor of the emitting oscillator
corresponding to its maximum coupling to the transmission line.
These requirements make this scheme impractical for a high-fidelity
($1-F \simeq 10^{-3}$) transfer between superconducting qubits or
microwave resonators, even though tunable couplers for
superconducting qubits have been demonstrated experimentally.
\cite{tunable,Bialczak} A search for practical schemes for such
transfer is currently under way.\cite{Gal-Geller,Cleland-comm}

    In this paper we consider a modification of the above scheme
with the primary difference being the use of tunable couplers for
both the emitting and receiving resonators. This drastically reduces
the required ON/OFF ratio and duration of the procedure: ON/OFF$\sim
1/\sqrt{1-F}$, $T\sim [Q_{\min}/\omega ]\ln (1-F)^{-1}$, where
$Q_{\min}\gg 1$ corresponds to the maximum available coupling of
both resonators with the transmission line. Actually, instead of
using the amplitude fidelity $F$, we will characterize the transfer
by the energy efficiency $\eta=F^2$. In the above formulas $F$ can
be replaced with $\eta$ because $1-F\approx (1-\eta)/2$. With these
improved requirements, we believe the information transfer between
superconducting qubits may become practical.

    As a tunable coupler we consider the experimentally realized
inductive coupler of Ref.\ \onlinecite{Bialczak}. In the analysis we
assume weak coupling (large $Q$-factors of the resonators) and slow
variation of the coupling compared to the resonator frequency. This
allows us to neglect field propagation within the resonators and
consider evolution of only one standing-wave mode in each resonator.
The main idea of the procedure is very simple: we tune the couplers
to cancel the back-reflection into the transmission line from the
receiving coupler.

   The next section is an overview of the work: we describe the
system and the procedure, derive main results in a simple
approximate way, and compare our system and results with those of
Ref.\ \onlinecite{Yurke}. In Section III we calculate the
$S$-parameters (transmission and reflection amplitudes) for a
tunable coupler of Ref.\ \onlinecite{Bialczak}. Section IV presents
a quantitative analysis of the procedure and estimates for the
technology of superconducting phase qubits. Section V is a
conclusion. Section III and a part of Sec.\ IV assume a particular
superconducting qubit system. Most of results of Secs.\ II and IV
are applicable to any system with realizable tunable couplers
between a resonator and a transmission line (hopefully, such tunable
couplers between an optical cavity and a fiber waveguide will be
developed in future).

\section{Overview: system, main idea, and approximate results}

    Our goal is to transfer a quantum state between two microwave
resonators via a transmission line (Fig.\ 1). This is done by
varying in time the coupling between the resonators and the
transmission line. Assuming sufficiently slow evolution (high-$Q$
resonators and slow change of the coupling) we consider only one
mode per resonator and assume the same frequency $\omega$ of these
modes (the effect of a frequency mismatch will be discussed in Sec.\
IV). Since in our simple case the quantum language essentially
coincides \cite{Yurke,Walls-Milburn,Yurke-84} with the classical
language (the classical field amplitudes should be associated with
annihilation operators in the Heisenberg picture), we will basically
discuss a classical field transfer between the resonators. The
resulting microwave phase in the receiving resonator (which
corresponds to a qubit phase) depends on the duration of propagation
through the transmission line. In our procedure we do not attempt to
control this phase and characterize the quality of the procedure by
the energy efficiency $\eta$. For simplicity we assume a lossless
and dispersionless transmission line and lossless resonators, so
that the relative energy loss $1-\eta$ is due to an imperfect
transmission/capture of the wave only.

\begin{figure}
\centering
\includegraphics[width=8.3cm]{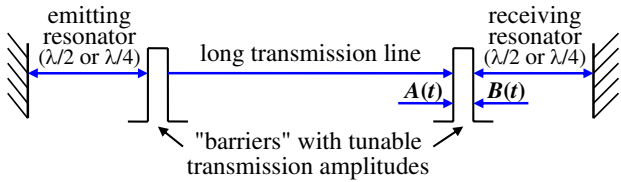}
%
\caption{Our goal is to pass a classical microwave energy from one
resonator to another one via a transmission line. Conceptually, this
is done by varying in time transmission amplitudes of ``barriers''
(tunable couplers) separating the resonators (half wavelength or
quarter wavelength in size) and the long transmission line. }
    \label{system}
    \end{figure}

    The {\it main idea} of our construction for a nearly perfect transfer
is the following. Suppose a microwave (voltage) waveform $A(t)\,
e^{i\omega t}$ is incident to the receiving coupler from the
transmission line (Fig.\ 1). The time dependence of the receiving
coupler parameters is chosen in the way, which eliminates the wave
reflected from the receiving coupler back into the transmission
line. This is done by arranging exact cancellation (destructive
interference) between the reflected wave and the transmitted part of
the wave $B(t)\, e^{i\omega t}$ from the receiving resonator. If the
reflection back into the transmission line is canceled, then all of
the microwave power is collected in the receiving resonator -- this
is exactly  our goal.
   Instead of varying in time the receiving coupling
to achieve a perfect cancellation of the refection, it is also
possible to keep it fixed, but vary the emitting coupling, designing
a specific $A(t)$ for a perfect cancellation. In a general case both
the emitting and receiving coupler parameters can be varied in time
in accordance with each other to satisfy just one equation:
cancellation of the reflection. Actually, this is a complex-number
equation because of an amplitude and a phase; however, we will see
later that in the case of equal resonator frequencies and weak
coupling the phase relation is satisfied automatically, so we are
left with only one real equation to be satisfied. In our particular
construction we vary only the emitting coupling in the first part of
the procedure, while keeping maximum the receiving coupling, and do
it in the  opposite way in the second part of the procedure: keep
the emitting coupling maximum and vary the receiving coupling. The
durations of the two parts are approximately equal.

    The perfect reflection cancellation is obviously impossible at the
beginning of the procedure, because it should take time to build up
the microwave amplitude $B(t)$ in the receiving resonator. The
microwave reflected back into the transmission line during the
build-up time is irrecoverably lost (it may actually lead to
multiple reflections, but we treat it as being lost -- see
discussion later). This means that we should design the waveform
envelope $A(t)$ to be very small during this initial period (see
Fig.\ 2), while after the build-up is finished, $A(t)$ can start
increasing rapidly. Let us choose it constant, $A(t)=A_0$, during
the build-up and crudely estimate the energy loss. Notice that we
shift the time origins at the emitter and receiver by the
propagation time, so that $t$ means both the emitting and receiving
time. Also notice that $A(t)$ can be assumed real.

\begin{figure}
\centering
\includegraphics[width=8cm]{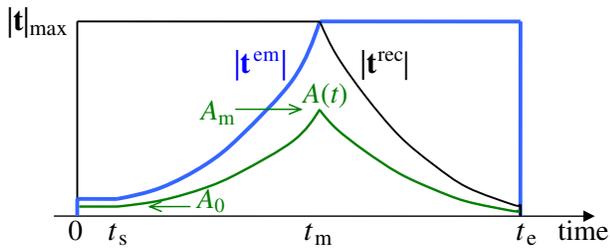}
%
\caption{(Color online) Sketches of the assumed time dependence
$|\t^{\rm em}(t)|$ of the transmission amplitude of the emitting
coupler (blue), transmission amplitude $|\t^{\rm rec}(t)|$ of the
receiving coupler (black), and the wave amplitude $A(t)$ incident to
the receiving coupler from the transmission line (green). The
voltage amplitude $A(t)$ is constant ($A_0$) during the build-up
period $0<t<t_{\rm s}$, then increases exponentially until time
$t_{\rm m}$ at which $|\t^{\rm em}|$ reaches the maximum available
value $|\t|_{\rm max}$, and after that $A(t)$ decreases
exponentially until the end of the process $t_{\rm e}$. The
receiving coupling $|\t^{\rm rec}|$ is kept at the maximum value
$|\t|_{\rm max}$ at $t<t_{\rm m}$, while at $t>t_{\rm m}$ the
emitting coupling $|\t|^{\rm em}$ is kept at the maximum $|\t|_{\rm
max}$. The ratio of couplings $|\t^{\rm em}/\t^{\rm rec}|$ is chosen
to cancel the back-reflection into the transmission line at any time
after $t_{\rm s}$.}
    \label{waveforms}
    \end{figure}

    During the build-up period the transmission amplitude $\t^{\rm rec}$
of the receiving coupler is kept at the maximum available value
$\t_{\rm max}$, which is still small, $|\t|_{\rm max}\ll 1$
(boldface is used for $\t$ to distinguish it from time; we will see
later that the phase of complex $\t$ is fixed). To start the perfect
reflection cancellation we need the wave amplitude $|B|$ in the
resonator to become $|A_0/\t_{\rm max}|$; this will happen at the
``start'' time $t_{\rm s} \sim \tau_{\rm bu}\equiv \tau_{\rm
rt}/|\t|_{\rm max}^2$, where $\tau_{\rm bu}$ is the time constant of
the build-up process, $t_{\rm s}\neq \tau_{\rm bu}$ because the
resonator leaks through the coupler, and $\tau_{\rm rt}$ is the
round-trip time of the wave in the resonator. For the
lowest-frequency mode
    \be
    \tau_{\rm rt}\approx 2\pi/\omega \hspace{0.3cm} {\rm or}
    \hspace{0.3cm} \tau_{\rm rt}\approx \pi/\omega
    \label{tau-rt-simple}\ee
for a $\lambda /2$ (half-wavelength) resonator or a $\lambda /4$
resonator, respectively (these equations are not exact because the
coupler may affect the boundary condition).
 Notice that our definition of the build-up time constant
$\tau_{\rm bu}$ has a direct relation to the resonator $Q$-factor
due to the coupler: in absence of the incoming wave the fraction
$|\t|^2$ of resonator energy would be lost every $\tau_{\rm rt}$,
therefore $Q=\omega \tau_{\rm rt}/|\t|^2$, and so $\tau_{\rm
bu}=Q_{\rm min}/\omega$ (in this derivation we implicitly assumed
the same wave impedance $R$ in the resonator and in the transmission
line). Since the travelling wave $A_0$ carries the power
$|A_0|^2/2R$, the energy loss during the build-up period $t_{\rm s}$
is $\sim |A_0|^2 \tau_{\rm bu}/R$, which should be later compared
with the total transmitted energy.

After $t_{\rm s}$ the transmitted energy is no longer being lost,
and $A(t)$ may increase as fast as $A_0 \exp [(t-t_{\rm
s})/2\tau_{\rm bu}]$. A simple way to derive this formula is by
applying time reversal to the ``no-reflection'' procedure, which
converts it into a ``no incident wave'' case, corresponding to a
leaking resonator. In such a case the wave amplitude in the
resonator decreases as $\exp (-\omega t/2Q)$ [the energy decreases
as $\exp (-\omega t/Q)$], and the leakage amplitude in the
transmission line has the same time dependence. Therefore for the
absence of reflection we need $A(t)\propto \exp (\omega t/2Q) $, and
for the assumed maximum coupling ($Q_{\rm min}=\omega \tau_{\rm
bu}$) we get $A(t)\propto \exp (t/2\tau_{\rm bu})$. The application
of time reversal also proves that the wave in the receiving
resonator automatically accumulates with a proper phase for the
reflection cancellation.

Exponentially increasing $A(t)$ requires a strong increase of the
emitting resonator coupling (slightly faster than exponential), so
this process can last only until some time $t_{\rm m}$, when the
emitting coupling reaches its physical limit (see Fig.\ 2). The
reflection cancellation can still be achieved after this time by
decreasing the receiving coupling. Assuming the same maximum
transmission $|\t|_{\rm max}$ and the same $\tau_{\rm rt}$ for the
emitting and receiving resonators, we find that after $t_{\rm m}$
the transmitted amplitude becomes exponentially decreasing,
$A(t)\propto \exp(-t/2\tau_{\rm bu})$, with the same time constant
as for the increasing part. To maintain the reflection cancellation,
the receiving coupling $\t^{\rm rec}$ should then be decreased in
time slightly faster than $\propto \exp(-t/2\tau_{\rm bu})$, since
the amplitude $B(t)$ in the receiving resonator continues to grow.

If we end the procedure at a finite time $t_{\rm e}$, then there
will be some energy left untransmitted/unreceived. From the
symmetry, if we choose $t_{\rm e}-t_{\rm m} \approx t_{\rm m}-t_{\rm
s}$, then this energy loss will be comparable to the energy loss
$\sim |A_0|^2 \tau_{\rm bu}/R$ during the build-up period $t_{\rm
s}$. Since the total transmitted energy is approximately $|A_0|^2
\exp [(t_{\rm m}-t_{\rm s})/\tau_{\rm bu}] \, \tau_{\rm bu}/R$, the
relative loss is $1-\eta \sim \exp [-(t_{\rm m}-t_{\rm s})/\tau_{\rm
bu}]$. Therefore, for an almost perfect efficiency $\eta$, the total
duration of the procedure, $t_{\rm e} \approx 2 (t_{\rm m}-t_{\rm
s})$, should be crudely
    \be
    t_{\rm e} \simeq 2\, \tau_{\rm bu} \ln \frac{1}{1-\eta}, \,\,\,
    \tau_{\rm bu}= \frac{\tau_{\rm rt}}{|\t|_{\rm max}^2} = \frac{
    Q_{\rm min}}{\omega} .
    \label{t-e-estimate}\ee

The required ON/OFF ratio $|\t|_{\rm max}/|\t|_{\rm min}$ for the
receiving coupler can be estimated as $\sqrt{2} A(t_{\rm
m})/A(t_{\rm e})=\sqrt{2} \exp [(t_{\rm e}-t_{\rm m})/2\tau_{\rm
bu}]$, where the factor $\sqrt{2}$ is needed because the energy in
the receiving resonator at time $t_{\rm m}$ is approximately half of
its energy at time $t_{\rm e}$. An estimate for the emitting coupler
is similar because $A(t_{\rm s})\sim A(t_{\rm e})$. Therefore, the
requirement is crudely
    \be
    {\rm ON/OFF} \simeq \frac{\sqrt{2}}{\sqrt{1-\eta}}.
    \label{ON/OFF-estimate}\ee

    Equations (\ref{t-e-estimate}) and (\ref{ON/OFF-estimate}) are
the main approximate results for our procedure; the exact results
will be presented in Sec.\ IV. A quick estimate using $|\t|_{\rm
max}^2\simeq 10^{-2}$, $\omega/2\pi = 6$ GHz, $\tau_{\rm rt}=
\pi/\omega$, and $\eta =0.999$, gives numbers quite encouraging for
an experiment with superconducting qubits: $t_{\rm e}\simeq 120$ ns
and ON/OFF $\simeq$ 45.

\vspace{0.5cm}

    Now let us discuss the differences
between our work and Ref.\ \onlinecite{Yurke}, and the reason why we
obtained so much better results for the duration $t_{\rm e}$ and
ON/OFF ratio.
   Most importantly, in Ref.\ \onlinecite{Yurke} only the emitting
coupling $\t^{\rm em}$ is modulated in time (receiving coupling
$\t^{\rm rec}$ is fixed), so that in the language of our Fig.\ 2 the
amplitude $A(t)$ continues to increase exponentially after $t_{\rm
m}$, until the emitting resonator is practically empty. Then, since
at $t_{\rm e}$ the ratio of wave amplitudes in the emitting and
receiving resonators should be $\sim \sqrt{1-\eta }$ (the energy
ratio is $\sim 1-\eta$), the ratio of transmissions is $\t^{\rm
rec}/\t^{\rm em}_{\rm max}\sim \sqrt{1-\eta}$. Similarly, at $t_{\rm
s}$ there should be the opposite relation of the wave amplitudes,
and therefore $\t^{\rm em}_{\rm min}/\t^{\rm rec}\sim
\sqrt{1-\eta}$. As a result, the required ON/OFF ratio is $\t^{\rm
em}_{\rm max}/\t^{\rm em}_{\rm min}\sim 1/(1-\eta )$, which is much
larger than our result (\ref{ON/OFF-estimate}). Since the receiving
resonator amplitude increases by $\sim 1/\sqrt{1-\eta}$ times
between $t_{\rm s}$ and $t_{\rm e}$, the duration of the procedure
can be estimated as $t_{\rm e}\simeq (Q^{\rm rec}/\omega ) \ln
(1-\eta)^{-1} \sim [Q^{\rm em}_{\rm min}/\omega (1-\eta )] \ln
(1-\eta)^{-1}$. This is $\sim 1/(1-\eta)$ times longer than our
result (\ref{t-e-estimate}) for the same $Q_{\rm min}^{\rm em}$.

    Actually, the requirements for the procedure duration and
ON/OFF ratio reported in Ref.\ \onlinecite{Yurke} are even stronger:
there are inequalities with ``much greater'' signs instead of the
signs ``comparable'' in our estimates. The strategy is also
different. The idea of reflection cancellation is not mentioned in
Ref.\ \onlinecite{Yurke} and our initial build-up period $t<t_{\rm
s}$ is absent. Instead, the result is obtained by a formal
optimization of the fidelity using the Euler-Lagrange equation. In
this case the reflection cancellation is not exact. However, this is
a relatively minor difference between Ref.\ \onlinecite{Yurke} and
our work. Even though the idea of reflection cancellation is not
discussed in Ref.\ \onlinecite{Yurke}, it is essentially there.

    As one more difference, the scheme of Ref.\  \onlinecite{Yurke}
assumes a circulator placed in between the resonators, so that the
back-reflected wave is diverted and fully absorbed. In our
derivation we essentially analyze the same situation treating the
reflected wave as a loss; however, we do it with a different
reasoning, not assuming a circulator. Formally, we can assume a very
long transmission line, so that the reflected wave does not play any
role. In a real experiment this is impossible, and the reflected
wave leads to multiple reflections from the two couplers, which
greatly complicates the analysis. However, there is a simple way to
show that the multiple reflections do not change the efficiency
$\eta$ significantly. Let us separate the wave amplitudes at time
$t_{\rm s}$ into two parts: the result of the simple analysis for
which $B(t_{\rm s})=-A_0/\t_{\rm max}$ and the remaining part. Then
using the system linearity we can analyze the further evolution of
both parts separately and add their contributions at the end time
$t_{\rm e}$. Since the reflection is fully cancelled in our usual
procedure after $t_{\rm s}$, the evolution of this part remains the
same as in the simple analysis. For the remaining part let us use
the energy conservation. If the length of the transmission line is
not resonant with the frequency $\omega$ (we assume this case), then
the energy of the remaining wave part cannot be larger than on the
order of $|A_0|^2 t_{\rm s}/R \sim E_{\rm in} (1-\eta)$, where
$E_{\rm in}$ is the initial energy of the emitting resonator. The
transfer efficiency is decreased if this wave gives an anti-phase
contribution at the receiving resonator at time $t_{\rm e}$. Instead
of considering the receiving resonator at $t_{\rm e}$, it is easier
to use conservation of the total energy and estimate the energy at
$t_{\rm e}$ in the emitting resonator and the transmission line.
Even if the two wave contributions (from simple analysis and due to
multiple reflections) are added there in-phase, the energy loss
cannot be larger than twice the sum of the corresponding energies,
so it is still on the order of $E_{\rm in} (1-\eta)$. Therefore even
in the worst-case scenario the inefficiency $1-\eta$ cannot increase
more than few times due to the effect of multiple reflections.
Moreover, such increase requires an unlucky phasing of the multiple
reflections, while on average $\eta$ is not decreased by their
contribution. This is why we neglect multiple reflections in our
analysis and consider the reflected wave as being lost.

    From the above analysis based on linearity and energy
conservation it is clear that the initial period $t_{\rm s}$ is
actually not needed. We can start the procedure of reflection
cancellation from $t=0$ just pretending that the needed amplitude
$A_0/\t_{\rm max}$ is already in the receiving resonator. The energy
loss at the start of the procedure will be then only few times
larger, and this can be compensated by a slight decrease of $A_0$.
The time dependence of $\t^{\rm em}$, $\t^{\rm rec}$, and $A$ on
Fig.\ 2 in this case will be fully symmetric about the middle point
$t_{\rm m}$. Even though the initial period $t_{\rm s}$ is not
necessary, we keep it in our analysis for conceptual simplicity.

    \section{$S$-parameters for inductive tunable coupler and
    $Q$-factor}

    The idea described in the previous section can be applied to
various systems which permit realization of tunable ``barriers''
between standing-wave modes in resonators and a propagating mode in
a transmission line. However, in this paper we will focus on a
particular technology developed for superconducting phase qubits,
\cite{Bialczak} which can be used to transfer microwave qubits
between two superconducting coplanar waveguide (or microstrip)
resonators. In particular, in this section we will calculate the
transmission and reflection amplitudes $\t$ and $\r$ for the tunable
coupler of Ref.\ \onlinecite{Bialczak}, assuming that this coupler
connects two coplanar waveguides (or microstrips). The readers
interested in other realization (optical, etc.) can skip this
section and go directly to Sec.\ IV, in which the transfer procedure
of Sec.\ II is analyzed in more detail.

Notice that the terminology of $S$-parameters (scattering matrix
elements) is the same in microwave technology and in quantum
mechanics. Instead of the notation $S_{11}$, $S_{12}$, etc., we use
a more transparent notation
$S=\left(\begin{array}{cc} \r_1& \t_2\\
\t_1&\r_2\end{array}\right)$, where the subscript indicates that the
incident wave came from the left (side 1) or from the right (side 2)
-- see Fig.\ 3. In the usual case of equal wave impedances from both
sides, $R_1=R_2=R$, the $S$-matrix should be unitary because of the
energy conservation, which means
$|\t_1|^2+|\r_1|^2=|\t_2|^2+|\r_2|^2=1$ and
$\r_2/\r_1^*=-\t_2/\t_1^*$ (so that necessarily $|\t_2|=|\t_1|$ and
$|\r_2|=|\r_1|$). The reciprocity condition in the usual case (same
for microwaves and in quantum mechanics) gives $\t_2=\t_1$ (so we
will use $\t$ without a subscript), and therefore the general form
is
    \be
    S=\left(\begin{array}{cc} \r_1& \t \\ \t& -\r_1^* \t/\t^*
    \end{array}\right) , \,\,\, |\t|^2+|\r_1|^2=1.
    \ee
Notice that if the coupler is symmetric, then $\r_1 =\r_2$, and this
is possible only if $\t$ and $\r$ are shifted by $\pm 90$ degrees;
in other words the ratio $\t/\r$ should be purely imaginary. In a
general asymmetric case $\t^2/\r_1\r_2$ is negative-real, and
therefore $\t/\sqrt{\r_1 \r_2}$ is purely imaginary.

\begin{figure}
\centering
\includegraphics[width=4.5cm]{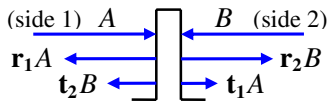}
%
\caption{Notations for the scattering matrix. }
    \label{scattering}
    \end{figure}

    Now let us consider the tunable coupler of Ref.\
\onlinecite{Bialczak} (for other ideas used for superconducting
qubits see Ref.\ \onlinecite{tunable}), and for simplicity
\cite{Pinto} let us describe it as an inductive coupler (Fig.\ 4)
with inductances $L_1$ and $L_2$, and mutual inductance $M$. In the
experiment \cite{Bialczak} $L_1$ and $L_2$ are practically constant,
while $M$ is tunable and small, $M^2\ll L_1 L_2$. The actual
circuit\cite{Bialczak} is more complex than in Fig.\ 4. The
effective coupling inductance $M$ consists of two contributions: a
fixed negative geometrical mutual inductance and a tunable positive
Josephson inductance of a shared Josephson junction, which is tuned
by controlling a bias current.
  In spite of simplicity, the model of Fig.\ 4 is a
good description, as confirmed by the experimental results
\cite{Bialczak} and quantum analysis. \cite{Pinto} Parasitic
elements like distributed stray capacitance surely make the model
more complicated; however, in the assumed weak-coupling limit their
possible effect is only a slight change of the transmission and
reflection amplitudes.

\begin{figure}
\centering
\includegraphics[width=8cm]{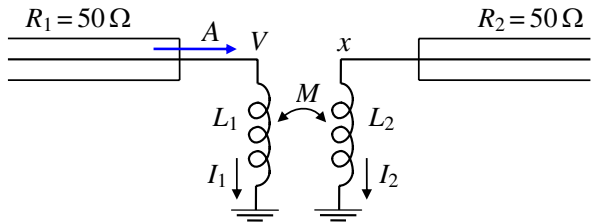}
%
\caption{The tunable coupler consisting of inductances $L_1$ and
$L_2$ coupled by a tunable mutual inductance $M$. Incoming wave
amplitude is $A$ (voltage), the voltages across the inductors are
$V$ and $x$, and the transmission amplitude is $\t=x/A$. }
    \end{figure}

    As a reminder, if a microwave transmission line is terminated with
a short, then $\r=-1$ (the voltage changes sign because the voltage
is zero at the short), while if it is terminated with a break, then
$\r=1$ (because then the current is zeroed). The short (very small
impedance) or break (very large impedance) should be compared with
the wave impedance of the transmission line $R\simeq 50\, \Omega$.
    We mostly consider the case of large inductances $L_{1,2}$ in the coupler,
$\omega L_{1,2}\gg R$, and weak coupling, $M^2\ll L_1 L_2$,  so that
$\r_1\approx \r_2 \approx 1$. The transmission amplitude $\t$ is
therefore almost purely imaginary. Let us first find $\t$ in this
high-impedance weak-coupling approximation, and then find it in a
general case.

    Assume that there is a voltage wave $Ae^{i\omega t}$ incident from the
left, and there is nothing incident from the right. Since
$\r_1\approx 1$, the voltage across $L_1$ is $V=2Ae^{i\omega t}$ (we
will often omit $e^{i\omega t}$ by using the phasor representation),
and the current in $L_1$ is $I_1=V/i\omega L_1$. It is important
that the voltage across $L_2$ is not $VM/L_1$ because $L_2$ is
loaded by a small resistance $R$. Let us denote the voltage across
$L_2$ as $x$; then the current in the outgoing transmission line is
$x/R$, and the current in $L_2$ is the same with opposite sign:
$I_2=-x/R$. Then from the standard equation for coupled inductors (a
transformer)
    \be
    x= M (i\omega I_1) +L_2 (i\omega I_2)= i\omega M \, \frac{V}{i\omega L_1} -  i\omega
    L_2 \, \frac{x}{R}
    \ee
we find $x=(VM/L_1)/(1+i\omega L_2/R)$. Since we assume $\omega
L_2\gg R$, this means $x=(VM/L_1) (R/i\omega L_2)$. Finally, using
the relation $V=2A$, we find the transmission amplitude $\t=x/A$ in
the high-impedance weak-coupling approximation:
    \be
  \t=-i \, \frac{2MR}{\omega L_1 L_2}, \,\,\,\, \r_2=\r_1
    =1-\frac{|\t|^2}{2},
    \label{t-simple}\ee
where the term $|\t|^2/2$ is included to satisfy $|\t|^2+|\r|^2=1$
in the leading order (actually, neglected imaginary contributions to
$\r_2$ and $\r_1$ can be larger than $|\t|^2$, but they are not so
important).

\vspace{0.5cm}

    Now let us consider a general case with arbitrary ratios
$\omega L_{1}/R_{1}$, $\omega L_2/R_2$, and $M^2/L_1L_2$, and
possibly different wave impedances $R_1$ and $R_2$ of the two
transmission lines. In quantum mechanics $R_1\neq R_2$ corresponds
to different potential energies or effective masses at the two
sides. In this case the scattering matrix $S$ is no longer unitary.
Instead, the matrix
$\tilde{S}=\left(\begin{array}{cc} \r_1& \t_2 \sqrt{R_2/R_1}\\
\t_1\sqrt{R_1/R_2}\,\, &\r_2\end{array}\right)$ is unitary; this
still follows from the conservation of energy. The reciprocity
condition is now $\t_1\sqrt{R_1/R_2}=\t_2 \sqrt{R_2/R_1}$, which
leads to the following general form of the $S$-matrix:
    \be
    S=\left(\begin{array}{cc} \r_1\,\, & \t_1 R_1/R_2 \\ \t_1& -\r_1^*
    \t_1/\t^*_1
    \end{array}\right) , \,\,\, |\t_1|^2\frac{R_1}{R_2}+|\r_1|^2=1.
    \ee
Notice that $|\r_1|=|\r_2|$ but $|\t_1|\neq |\t_2|$.

    The derivation for $\t_1$ and $\r_1$ is now slightly different
because the voltage $V=A(1+\r_1)$ is unknown, so we have a system of
two equations:
    \begin{eqnarray}
&& x = i\omega M \frac{(1-\r_1)A}{R_1}-i\omega L_2 \frac{x}{R_2},
  \label{t-der-1}\\
&&  (1+\r_1) A =i\omega L_1 \frac{(1-\r_1)A}{R_1} -i\omega M
  \frac{x}{R_2}
    \label{t-der-2}\end{eqnarray}
(notice that now $I_1 \neq V/i\omega L_1$), from which we find
$\t_1=x/A$ and $\r_1$:

    \begin{eqnarray}
&& \hspace{-0.6cm} \t_1 = -i \, \frac{2 MR_2}{\omega (L_1L_2-M^2)}
    \nonumber \\
&& \hspace {-0.2cm}\times \left[ 1-i\frac{R_2 L_1+R_1 L_2}{\omega
(L_1L_2-M^2)}-\frac{R_1R_2}{\omega^2 (L_1L_2-M^2)} \right]^{-1},
\,\,\,
    \label{t-gen-m}  \\
&& \hspace{-0.6cm}   \r_1 =  \frac{ \displaystyle{ 1-i\frac{R_2
L_1-R_1 L_2}{\omega (L_1L_2-M^2)}+\frac{R_1R_2}{\omega^2
(L_1L_2-M^2)}}}{ \displaystyle{ 1-i\frac{R_2 L_1+R_1 L_2}{\omega
(L_1L_2-M^2)}-\frac{R_1 R_2}{\omega^2 (L_1L_2-M^2)}} } \, .
  \label{r_l-gen-m}  \end{eqnarray}

It is easy to check the relation $|\t_1|^2 (R_1/R_2)+|\r_1|^2=1$
explicitly. The $S$-parameters $\t_2=\t_1 R_1/R_2$ and
$\r_2=-\r_1^*\t_1/\t_1^*$ are given by Eqs.\ (\ref{t-gen-m}) and
(\ref{r_l-gen-m}) with exchanged parameters $R_1\leftrightarrow
R_2$, $L_1\leftrightarrow L_2$.
 It is easy to see that
$\t_1/(\sqrt{\r_1}\sqrt{\r_2})$ is negative-imaginary if the square
root is defined in the natural way and $M$ is assumed to be positive
for definiteness.

    An important limiting case of Eqs.\ (\ref{t-gen-m}) and
(\ref{r_l-gen-m}) is when $M^2/L_1L_2 \ll 1 $ and
$R_1R_2/\omega^2L_1L_2$ is not too close to 1. Then the coupling is
weak, while the impedance ratios $\omega L_1/R_1$ and $\omega
L_2/R_2$ are arbitrary. In this case
    \begin{eqnarray}
&& \hspace{0.0cm} \t_1 = \frac{2i \omega MR_2}{(i\omega L_1
+R_1)(i\omega L_2+R_2)}
 \,, \,\,\, \t_2 =\t_1 \frac{R_1}{R_2}\, , \qquad
    \label{t-lim-m}  \\
&& \hspace{0.0cm}   \r_1 = \frac{i\omega L_1-R_1}{i\omega L_1+R_1}
\,  \left(1-\frac{|\t_1|^2}{2}\frac{R_1}{R_2} \right) , \,\,
    \\
&& \hspace{0.0cm}  \r_2 = \frac{i\omega L_2-R_2}{i\omega L_2+R_2}
\left(1-\frac{|\t_1|^2}{2}\frac{R_1}{R_2} \right) ,
  \label{r-lim-m}  \end{eqnarray}
where the factors $1-|\t_1|^2(R_1/R_2)/2$ are included again to
satisfy the energy conservation in the leading order.

\vspace{0.5cm}

If a wave $A$ is incident from the left, then the transmitted power
is $P=|A\t_1|^2/2R_2$. It is useful to express this power in terms
of the voltage $V=A(1+\r_1)$ across $L_1$. In the weak coupling case
we find the ratio
 \be
 \frac{P}{|V|^2}=\frac{|\t_1|^2}{2R_2|1+\r_1|^2}=
 \frac{M^2 R_2}{2\omega^2 L_1^2 L_2^2 [1+(R_2/\omega L_2)^2]},
 \ee
which obviously does not depend on $R_1$.

    This formula can be used to find the $Q$-factor ($Q\gg 1$) of a
resonator or a lumped-circuit oscillator of frequency $\omega$
connected to the coupler from the left. The energy $E$ of an
oscillator decreases because of the leaking power $P$, and the
corresponding $Q=\omega E/P$ is
    \be
    Q= \frac{E}{|V|^2} \,
    \frac{2\omega^3 L_1^2 L_2^2 [1+(R_2/\omega L_2)^2]}{M^2 R_2},
    \label{Q-viaV}\ee
where the ratio $E/|V|^2$ depends on the oscillator type and
parameters.

    For example, for a coplanar resonator, in which the energy
is mainly stored ``inside'', with negligible contributions from the
lumped elements at the two edges (then $\omega L_1\gg R_1$,
$\r_1\approx 1$), we have
         \be
V = 2 A, \,\, E =\frac{|A|^2}{2R_1}\tau_{\rm rt} = l C_{\rm pl}
|A|^2,
    \ee
where $l$ is the resonator length, $C_{\rm pl}$ is the capacitance
per unit length, the round-trip time is $\tau_{\rm rt}=2l/v$, and
the wave velocity is $v=1/C_{\rm pl}R_1$ (as follows from the
telegraph equations). This case corresponds to Eq.\
(\ref{tau-rt-simple}).

 If the resonator energy has a significant
contribution from the lumped elements at the edges, then it is
meaningful to {\it define} the round-trip time $\tau_{\rm rt}$ via
the relation between the resonator energy $E$ and the ``traveling''
power $|A|^2/2R_1$, so that
    \be
   \tau_{\rm rt} \equiv \frac{2R_1E}{|A|^2}, \,\,\, \frac{E}{|V|^2}=
    \frac{\tau_{\rm rt} [1+(R_1/\omega L_1)^2]}{8R_1} ,
    \label{tau-rt-def}\ee
where we used $V/A=1+\r_1$. In this case
    \begin{eqnarray} &&
\hspace{-0.5cm} Q= \frac{\tau_{\rm rt}\omega^3 L_1^2 L_2^2
[1+(R_1/\omega L_1)^2] [1+(R_2/\omega L_2)^2]}
    {4M^2 R_1 R_2}  \quad
    \label{Q-tau-rt-1}\\
&& = \frac{\omega \tau_{\rm rt}}{|\t|_1^2}\frac{R_2}{R_1}   \, ,
    \label{Q-tau-rt-2}\end{eqnarray}
where the last formula can also be easily derived directly as
$Q=\omega E/P$.

 If we use a lumped-element $LC$-oscillator, for
example a superconducting phase qubit with effective parameters
$L_{\rm qb}$ and $C_{\rm qb}$, then
    \be
    \frac{E}{|V|^2} = \frac{C_{\rm qb}}{2}, \,\,\,
 Q=   \frac{C_{\rm qb}\omega^3 L_1^2 L_2^2 [1+(R_2/\omega L_2)^2]}{M^2
 R_2}  \, .
    \label{Q-qubit}\ee

    \section{Quantitative analysis of the procedure and estimates}

    Let us analyze quantitatively the procedure of the wave
 transfer discussed in Sec.\ II.
   Assume that a voltage wave $A(t)\, e^{i\omega t}$ is incident from the
transmission line (from the left, which is side 1 in our notation)
to the tunable coupler of the receiving resonator (Fig.\ 1). The
wave incident from the resonator side (side 2) is $B(t)\, e^{i\omega
t}$, and we again assume the same wave impedance from both sides,
$R_1=R_2=R\simeq 50\, \Omega$. As discussed in Sec.\ II, the idea is
to cancel the wave emitted back into the transmission line. Exact
cancellation occurs if
    \be
    \r_1 A +\t  B =0,
    \label{cancel}\ee
so we should vary the receiving coupler parameters and/or $A(t)$ to
satisfy this equation at any time except for the initial period
$t<t_{\rm s}$. Here $\t = \t^{\rm rec}$,  $\r_1 = \r_1^{\rm rec}$;
for brevity of notations we will mostly omit the superscript for the
receiving coupler parameters, while using the superscript ``em'' for
the emitting coupler.

    The evolution of the receiving resonator amplitude $B(t)$
in the high-$Q$ small-detuning case can be described as
\cite{note-evol}
    \be
    \dot{B}= \frac{\r_2 e^{i\varphi}-1}{\tau_{\rm rt}} B +
   \frac{e^{i\varphi}}{\tau_{\rm rt}}  A\t .
    \label{B-evol}\ee
Here $\tau_{\rm rt}$ is the round-trip time for the wave in the
resonator [see Eq.\ (\ref{tau-rt-simple})], which can be defined in
general as  $\tau_{\rm rt}\equiv 2R E/|B|^2$, where $E$ is the
resonator energy [see Eq.\ (\ref{tau-rt-def})]. The parameter
$\varphi$ describes deviation of the effective resonator length from
the perfect value, so that the detuning $\delta\omega=\omega_{\rm
rec}-\omega$ of the resonator frequency is $\delta \omega = (\varphi
+\arg \r_2) /\tau_{\rm rt}$. [If $\r_2\approx -1$, then $\varphi
\approx \pi$; however, here we focus on the case when $r_2\approx
1$, which requires $\varphi\approx 0$.] Non-zero $\varphi$ is needed
if the ratio $\omega L_{2}/R$ is finite (then $\arg \r_2 \neq 0$)
and we want $\delta\omega =0$. The detuning can in principle be
varied in time;\cite{Chalmers} we do not explicitly consider this
possibility below, but our assumption $\delta\omega =0$ actually
implies slight variation of $\varphi$, because $\arg \r_2$ slightly
changes when $\t$ is modulated. The resonator $Q$-factor is
$Q=\omega\tau_{\rm rt}/|\t|^2$, which follows from Eq.\
(\ref{B-evol}) when $Q\gg 1$ and $|\delta\omega | \ll 1/\tau_{\rm
rt}$, since $\mbox{Re}(1-\r_2 e^{i\varphi})\approx 1-|\r_2|\approx
|\t|^2/2$.

    The easiest case to analyze is the weak-coupling high-impedance
limit for the receiving coupler, when $|M|\ll L_1L_2$ and $\omega
L_{1,2}\gg R$ (as discussed later, the high-impedance assumption is
actually not important). Then $\r_1 =\r_2 = 1-|\t|^2/2\approx 1$,
$\t$ is negative-imaginary [see Eq.\ (\ref{t-simple}); we assume
$M>0$ for definiteness], and we also assume $\varphi =0$ to get
$\delta\omega =0$. (Actually, as discussed above, $\varphi$ should
slightly vary in time to compensate the frequency detuning due to
the coupler, since we assume $\delta\omega =0$.) In this case Eq.\
(\ref{B-evol}) becomes
   \be
    \dot{B}= -\frac{|\t|^2}{2\tau_{\rm rt}} B +
    \frac{\t}{\tau_{\rm rt}} A .
    \label{B-evol-2}\ee
Notice that $A(t)$ comes from the emitting resonator having the same
frequency, and it comes through a weak coupler also; this is why its
phase does not change with time, and we can assume $A(t)$ to be real
and positive, $A(t)>0$, by a trivial phase shift. Since $A(t)>0$ and
$\t$ is negative-imaginary, $B(t)$ (which starts from zero) is also
negative-imaginary.

    As was discussed in Sec.\ II, we keep
the transmitted amplitude constant, $A(t)=A_0>0$ for the build-up
period (Fig.\ 2), and keep $\t$ at its maximum value $\t=-i
|\t|_{\rm max}$. Solution of Eq.\ (\ref{B-evol-2}) which starts with
$B(0)=0$ is then
    \be
    B(t)=\frac{-2iA_0}{|\t|_{\rm max}} \,  (1-e^{-t/2\tau_{\rm bu}}) ,
    \ee
where
    \be
\tau_{\rm bu}=\frac{\tau_{\rm rt}}{|\t|_{\rm max}^2} =\frac{Q_{\rm
min}}{\omega}
    \label{t_bu}\ee
is the time constant of the build-up process, corresponding to the
maximum available coupling (minimum available $Q$-factor) of the
receiving resonator. The full reflection compensation can be started
when we reach $i|\t|_{\rm max} B=A_0$, which happens at time
    \be
    t_{\rm s}= \tau_{\rm bu} \ln 4.
    \label{t-s}\ee

    After that the reflection is kept cancelled by satisfying Eq.\
(\ref{cancel}); therefore
    \be
    \t=-\frac{A}{B}, \,\,\, \dot{B}=
        -\frac{A^2}{B}\frac{1}{2\tau_{\rm rt}} .
    \label{evol-perfect}\ee
Solution of this equation for $B(t)$ corresponds to the energy
conservation:
    \be
  B(t) = - i \sqrt { \left( \frac{A_0}{|\t|_{\rm max}}\right)^2
 +  \frac{1}{\tau_{\rm rt}} \int_{t_{\rm s}}^t A^2(t')\, dt'},
    \label{B(t)}\ee
where $A(t)$ at $t>t_{\rm s}$ is in general arbitrary, just with the
limitation $A(t)\leq |\t|_{\rm max} |B|$ (notice that since we can
vary $\t$, the ratio $A/iB$ may vary in time). Choosing an increase
of $A(t)$ at the maximum of this limitation [i.e.\ assuming $|\t
(t)|=|\t|_{\rm max}$] we obtain [see Eq.\ (\ref{evol-perfect})]
    \be
    A(t) = A_0 e^{(t-t_{\rm s})/2\tau_{\rm bu}}, \,\,\,
B(t) = \frac{-i A (t)}{|\t|_{\rm max}} , \,\,\, t\geq t_{\rm s}.
    \label{evol-increase}\ee

    The exponential increase (\ref{evol-increase}) of $A(t)$ requires
an increase of the transmission amplitude $|\t^{\rm em}|$ of the
emitting coupler. Assuming that at time $t_{\rm m}$ it reaches a
physical limitation $|\t|_{\rm max}^{\rm em}$ and is then kept
constant at this maximum available level, we obtain the change from
the form of Eq.\ (\ref{evol-increase}) to
    \be
    A(t)= A_{\rm m} e^{-(t-t_{\rm m})/2\tau_{\rm bu}^{\rm em}}  , \,\,\,
   |\t^{\rm rec} (t)| = \frac{-iA(t)}{B(t)}, \,\,\, t\geq t_{\rm m},
    \label{evol-decrease}\ee
where  $A_{\rm m}=A_0 e^{(t_{\rm m}-t_{\rm s})/2\tau_{\rm bu}}$ is
the maximum wave amplitude, $B(t)$ is given by Eq.\ (\ref{B(t)}),
and $\tau_{\rm bu}^{\rm em} \equiv \tau_{\rm rt}^{\rm em}/(|\t|_{\rm
max}^{\rm em})^2= Q_{\rm min}^{\rm em}/\omega$. Even though there is
no ``build-up'' at the emitting resonator, we use notation
$\tau_{\rm bu}^{\rm em}$ because the decay time constant is the same
as the build-up time constant, as is obvious from the time reversal
discussed in Sec.\ II.

    Let us assume that the procedure ends abruptly at time $t_{\rm e}$
and calculate the total energy loss. Since the energy carried by a
travelling wave $A(t)$ is $\int |A|^2 dt/2R$, the untransmitted
energy is $A_{\rm m}^2 e^{-(t_{\rm e}-t_{\rm m})/\tau_{\rm bu}^{\rm
em}} \tau_{\rm bu}^{\rm em}/2R$. The energy loss during the initial
build-up time $t_{\rm s}$ is $[A_0^2 t_{\rm s}- (A_0/|\t|_{\rm
max})^2\tau_{\rm rt}]/2R=A_0^2\tau_{\rm bu}(\ln 4 -1)/2R$. Therefore
the total relative loss is
    \be
   1-\eta = \frac{A_0^2 \tau_{\rm bu} (\ln 4 -1)
+A_{\rm m}^2 e^{-(t_{\rm e}-t_{\rm m}) /\tau_{\rm bu}^{\rm em}}
\tau_{\rm bu}^{\rm em}}
 {A_0^2 \tau_{\rm bu} (\ln 4 -1)+A_{\rm m}^2 \tau_{\rm bu}
  +A_{\rm m}^2  \tau_{\rm bu}^{\rm em}} ,
    \label{1-eta}\ee
where the denominator corresponds to the total (initially stored)
energy. Neglecting the term $A_0^2\tau_{\rm bu}(\ln 4 -1)$ in the
denominator (i.e.\ assuming $e^{(t_{\rm m}-t_{\rm s})/\tau_{\rm
bu}}\gg 1$) and using $A_{\rm m}=A_0 e^{(t_{\rm m}-t_{\rm
s})/2\tau_{\rm bu}}$, we obtain
    \be
1-\eta = \frac{(\ln 4 -1)e^{-(t_{\rm m}-t_{\rm s}) /\tau_{\rm
bu}^{\rm rec}}\tau_{\rm bu}^{\rm rec}+e^{-(t_{\rm e}-t_{\rm m})
/\tau_{\rm bu}^{\rm em}}\tau_{\rm bu}^{\rm em} }{\tau_{\rm bu}^{\rm
rec}+\tau_{\rm bu}^{\rm em}},
    \label{1-eta-2}    \ee
where the notation $\tau_{\rm bu}^{\rm rec}\equiv \tau_{\rm bu}$ is
used for symmetry.
    Notice that even though in the derivation we used the terminology
applicable to a resonator (for example, the round-trip time), the
result (\ref{1-eta-2}) for the efficiency $\eta$ depends only on the
minimum $Q$-factors of the resonators. If a resonator is replaced by
a phase qubit, then the only difference is a different formula for
the $Q$-factor [see Eqs.\ (\ref{Q-viaV}), (\ref{Q-tau-rt-1}), and
(\ref{Q-qubit})].

 To find the shortest duration of the procedure $t_{\rm e}$ for a fixed
$\eta$, we minimize this expression over $t_{\rm m}$, that gives
    \be
t_{\rm e} =  (\tau_{\rm bu}^{\rm em}+ \tau_{\rm bu}^{\rm rec})   \ln
\frac{1}{1-\eta}  + \tau_{\rm bu}^{\rm rec}   \ln [4(\ln 4 -1)].
 \label{duration-opt-1}\ee
In particular, in the symmetric case when $\tau_{\rm bu}^{\rm em}=
\tau_{\rm bu}^{\rm rec}=\tau_{\rm bu}$, the minimum duration is
    \be
 t_{\rm e} = 2 \tau_{\rm bu} \ln \frac{2\sqrt{\ln
4 -1}}{1-\eta} = \frac{2Q_{\rm min}}{\omega} \ln \frac{2\sqrt{\ln 4
-1}}{1-\eta}  ,
    \label{duration-opt}\ee
which is quite close to the estimate (\ref{t-e-estimate}), since
$\sqrt{\ln 4 -1}\approx 0.62$.

At the minimum duration (\ref{duration-opt-1}) the ratio of the
energy losses at the build-up and at the end is $\tau_{\rm bu}^{\rm
rec}/\tau_{\rm bu}^{\rm em}$, and the relations of the wave
amplitudes are $A_{\rm m}/A_0=\sqrt{\ln 4 -1}/\sqrt{1-\eta}$ and
$A_{\rm m}/A_{\rm e}=1/\sqrt{1-\eta}$ [see Eq.\ (\ref{1-eta-2})],
where $A_{\rm e}=A_{\rm m}e^{-(t_{\rm e}-t_{\rm m})/2\tau_{\rm
bu}^{\rm em}}$ is the wave amplitude at the end of the procedure.
The ratio of the transmitted energies before and after $t_{\rm m}$
is $\tau_{\rm bu}^{\rm rec}/\tau_{\rm bu}^{\rm em}$ [see denominator
of Eq.\ (\ref{1-eta})]; notice that the corresponding ratio of
energies in the receiving and emitting resonators at $t_{\rm m}$ is
$Q_{\rm min}^{\rm rec}/Q_{\rm min}^{\rm em}$, so that the same wave
amplitudes are emitted into the transmission line.

The maximum/minimum ratios for the transmission $\t$ of the emitting
and receiving couplers can be easily found from the calculated
ratios $A_{\rm m}/A_0$ and $A_{\rm m}/A_{\rm e}$, and the energy in
each resonator at $t_{\rm m}$; this gives
    \begin{eqnarray}
&& \frac{\t^{\rm em}_{\rm max}}{\t^{\rm em}_{\rm min}} =
\sqrt{\frac{Q^{\rm em}_{\rm min}}{Q^{\rm em}_{\rm max}}} =
\frac{\sqrt{\ln 4 -1}\sqrt{1+\tau_{\rm bu}^{\rm rec}/\tau_{\rm
bu}^{\rm em}}}{\sqrt{1-\eta}}, \qquad
    \label{ON/OFF-em}    \\
&&  \frac{\t^{\rm rec}_{\rm max}}{\t^{\rm rec}_{\rm min}} =
\sqrt{\frac{Q^{\rm rec}_{\rm min}}{Q^{\rm rec}_{\rm max}}} =
 \frac{\sqrt{1+\tau_{\rm bu}^{\rm em}/\tau_{\rm
bu}^{\rm rec}}}{\sqrt{1-\eta}}
    \label{ON/OFF}\end{eqnarray}
These are the ON/OFF ratios for the tunable couplers. In the
symmetric case when $\tau_{\rm bu}^{\rm em}=\tau_{\rm bu}^{\rm rec}$
they are
    \be
\frac{\t^{\rm em}_{\rm max}}{\t^{\rm em}_{\rm min}} =
\frac{\sqrt{2(\ln 4 -1)}}{\sqrt{1-\eta}}, \,\,\,
 \frac{\t^{\rm rec}_{\rm
max}}{\t^{\rm rec}_{\rm min}} = \frac{\sqrt{2}}{\sqrt{1-\eta}},
    \label{ON/OFF-sym}\ee
so that the required ON/OFF ratio is larger for the receiving
coupler, and it exactly coincides with the estimate
(\ref{ON/OFF-estimate}). In particular, for $\eta =0.999$ the ON/OFF
ratio is 28 for the emitting coupler and 45 for the receiving
coupler.

    For the requirements on the ON/OFF ratios of the couplers, it is
instructive to calculate the relative energy loss at the build-up
period and at the end of the process in the following way. If at the
build-up period we use $\t_{\rm min}^{\rm em}$ for the emitting
coupler (constant $A_0$ implies practically constant $\t^{\rm em}$)
and $\t_{\rm max}^{\rm rec}$ for the receiving coupler, then the
energy loss is $A_0^2 (\ln 4-1)\tau_{\rm rt}^{\rm rec}/|\t_{\rm
max}^{\rm rec}|^2 2R$, while the energy stored in the emitting
resonator is $(A_0/|\t_{\rm min}^{\rm em}|)^2\tau_{\rm rt}^{\rm
em}/2R$. Therefore the relative energy loss during the build-up is
 $1-\eta_{\rm bu} = (\ln 4 -1) (|\t|_{\rm min}^{\rm em}/|\t|_{\rm max}^{\rm rec})^2
(\tau_{\rm rt}^{\rm rec}/\tau_{\rm rt}^{\rm em})$ independently of
what happens later. Similarly, if at the end of the procedure we use
$\t_{\rm max}^{\rm em}$ for the emitting coupler and stop the
procedure when the receiving coupling needs to go below $\t_{\rm
min}^{\rm rec}$, then the relative energy loss is $1-\eta_{\rm end}
= (|\t|_{\rm min}^{\rm rec}/|\t|_{\rm max}^{\rm em})^2 (\tau_{\rm
rt}^{\rm em}/\tau_{\rm rt}^{\rm rec})$. Therefore the minimized
ON/OFF ratios for both couplers for a fixed $\eta =\eta_{\rm
bu}+\eta_{\rm end}-1$ are
    \be
    \frac{\t_{\rm max}^{\rm em}}{\t_{\rm min}^{\rm em}}=
    \frac{\t_{\rm max}^{\rm rec}}{\t_{\rm min}^{\rm rec}}=
    \frac{\sqrt{ 2\sqrt{\ln 4 -1}}}{\sqrt{1-\eta}}.
    \label{ON/OFF-min}\ee
This optimization differs from minimization of the duration $t_{\rm
e}$ and assumes a variable ratio $|\t|_{\rm max}^{\rm em}/|\t|_{\rm
max}^{\rm rec}$. However, we see that the ON/OFF ratio
(\ref{ON/OFF-sym}) for $\t^{\rm rec}$ obtained in optimizing $t_{\rm
e}$ in the case $\tau_{\rm bu}^{\rm em}=\tau_{\rm bu}^{\rm rec}$ is
only 30\% larger than the minimum (\ref{ON/OFF-min}).

    So far all formulas in this section starting with Eq.\
(\ref{B-evol-2}) were based on  the high-impedance weak-coupling
assumption, so that Eq.\ (\ref{t-simple}) can be used. However, the
results are essentially the same in the case when impedances of the
couplers are arbitrary, while the coupling is still weak, so that we
can use Eqs.\ (\ref{t-lim-m})-(\ref{r-lim-m}). This is because the
phases of $\t$, $\r_1$, and $\r_2$ do not change when the coupling
is tuned. Since the effective frequencies of the emitting and
receiving resonators are equal, then $\varphi = -\arg \r_2$ in Eq.\
(\ref{B-evol}), and therefore $\arg (B/A)=\arg (\t/\r_2)$. It is
easy to see that this automatically leads to the proper phase for
the reflection cancellation in Eq.\ (\ref{cancel}), since $\arg (-\t
B/A\r_1)=\arg (-\t^2/\r_1\r_2)=0$. Therefore, all formulas in this
section remain valid, just with some fixed phase shifts for $A(t)$
and $B(t)$.

    Equation (\ref{1-eta-2}) can also be used to analyze the case
when the receiving coupler is not controlled, as in Ref.\
\onlinecite{Yurke}. Then $t_{\rm e}=t_{\rm m}$, and from the second
term in the numerator we obtain $\tau_{\rm bu}^{\rm rec}>\tau_{\rm
bu}^{\rm em}/(1-\eta)$, assuming $1-\eta \ll 1$. For the duration we
use the first term and find $t_{\rm e}> (t_{\rm m}-t_{\rm
s})>\tau_{\rm bu}^{\rm rec} \ln [(\ln 4-1)/(1-\eta)]$; then using
the first inequality we obtain $t_{\rm e}> [\tau_{\rm bu}^{\rm
em}/(1-\eta)] \ln [(\ln 4-1)/(1-\eta)]$. The minimized value of
$t_{\rm e}$ is close to this bound, but even its approximate formula
is rather lengthy: $t_{\rm e} \approx [\tau_{\rm bu}^{\rm
em}(1+y)/(1-\eta)] \ln [4(\ln 4-1)/y(1-\eta)] $, where
$y=1/\ln[4(\ln 4-1)/[(1-\eta)\ln (1-\eta)^{-1}]]$.
 For the ON/OFF ratio we notice that $A^2(t)$ should
increase during the procedure by the factor $e^{-(t_{\rm e}-t_{\rm
m}) /\tau_{\rm bu}^{\rm rec}}>(\ln 4 -1)/(1-\eta)$, while the energy
of the receiving resonator should decrease by more than $1/(1-\eta)$
times. Therefore, the ON/OFF ratio for the emitting coupler should
exceed $\sqrt{\ln 4 -1}/(1-\eta)$. The minimized ON/OFF ratio can be
shown to be twice larger. Comparing these results with Eqs.\
(\ref{duration-opt}) and (\ref{ON/OFF-sym}), we see that additional
use of a tunable receiving coupler shortens the procedure crudely by
the factor $1/2(1-\eta)$ assuming the same $\tau_{\rm bu}^{\rm em}$,
while the ON/OFF ratio is decreased crudely by the factor
$1/\sqrt{1-\eta}$.

    As mentioned at the end of Sec.\ II, the initial period $t_{\rm
s}$ is not really needed in our procedure. Let us use $t_{\rm s}=0$
and start the reflection cancellation (\ref{evol-increase}) just
pretending than the needed amplitude $B=-A_0/\t_{\rm max}$ is
already in the receiving resonator at $t=0$. Then using linearity we
see that for such procedure the initial loss of energy into the
transmission line is $A_0^2\tau_{\rm bu}^{\rm rec}/2R$, which is 2.6
times larger than the initial loss $A_0^2\tau_{\rm bu}^{\rm rec}(\ln
4 -1)/2R$ in our usual procedure. Then Eq.\ (\ref{1-eta-2}) is
replaced with
    \be
1-\eta = \frac{e^{-t_{\rm m} /\tau_{\rm bu}^{\rm rec}}\tau_{\rm
bu}^{\rm rec}+e^{-(t_{\rm e}-t_{\rm m}) /\tau_{\rm bu}^{\rm
em}}\tau_{\rm bu}^{\rm em} }{\tau_{\rm bu}^{\rm rec}+\tau_{\rm
bu}^{\rm em}},
    \label{1-eta-3}    \ee
and minimization of time $t_{\rm e}$ for a given $\eta$ is achieved
when $t_{\rm m}/\tau_{\rm bu}^{\rm rec}=(t_{\rm e}-t_{\rm
m})/\tau_{\rm bu}^{\rm em}$, so that $A_{\rm m}/A_{\rm 0}=A_{\rm
m}/A_{\rm e}=1/\sqrt{1-\eta}$. The total duration of such procedure
is
    \be
t_{\rm e} =  (\tau_{\rm bu}^{\rm em}+ \tau_{\rm bu}^{\rm rec})   \ln
\frac{1}{1-\eta}
 \label{duration-opt-3}\ee
instead of Eq.\ (\ref{duration-opt-1}), and the emitting coupler
ON/OFF ratio is
    \be
    \t^{\rm em}_{\rm max}/\t^{\rm em}_{\rm min} =
\sqrt{1+\tau_{\rm bu}^{\rm rec}/\tau_{\rm bu}^{\rm
em}}/\sqrt{1-\eta}
    \ee
 instead of Eq.\ (\ref{ON/OFF-em}), while for the
receiving coupler Eq.\ (\ref{ON/OFF}) is still valid. Therefore,
cutting out the initial period $t_{\rm s}$ makes the procedure
slightly shorter, and it makes it fully symmetric in time for a
symmetric system. Explicitly, in this ``pretending'' procedure the
tunable couplers should be varied in the following way (we still use
high-$Q$ approximation):
    \begin{eqnarray}
&& \hspace{-0.3cm} \t^{\rm em}(t) =  \frac{ \t^{\rm em}_{\rm max}\,
\sqrt{\tau_{\rm bu}^{\rm em}/\tau_{\rm bu}^{\rm rec}} }{
\sqrt{(1+\frac{\tau_{\rm bu}^{\rm em}}{\tau_{\rm bu}^{\rm rec}}) \,
e^{(t_{\rm m}-t)/\tau_{\rm bu}^{\rm rec}}-1 } }, \,\,0<t<t_{\rm m},
\qquad
    \label{t-em}\\
&& \hspace{-0.3cm} \t^{\rm rec}(t) = \frac{ \t^{\rm rec}_{\rm max}\,
\sqrt{\tau_{\rm bu}^{\rm rec}/\tau_{\rm bu}^{\rm em}} }{
\sqrt{(1+\frac{\tau_{\rm bu}^{\rm rec}}{\tau_{\rm bu}^{\rm em}}) \,
e^{(t-t_{\rm m})/\tau_{\rm bu}^{\rm em}}-1 } }, \,\, t_{\rm
m}<t<t_{\rm e}, \qquad
    \label{t-rec}\\
&& \hspace{-0.3cm}      t_{\rm m}=\tau_{\rm bu}^{\rm
rec}\ln\frac{1}{1-\eta} , \,\,\,  t_{\rm e}-t_{\rm m}=\tau_{\rm
bu}^{\rm em}\ln\frac{1}{1-\eta} . \qquad
    \label{tmte}\end{eqnarray}
Notice that the time dependence in Eq.\ (\ref{t-em}) is essentially
the same as in Ref.\ \onlinecite{Yurke}. [For the procedure with the
build-up period $t_{\rm s}$ the corresponding time-dependences
$\t^{\rm em}(t)$ and $\t^{\rm rec} (t)$ are similar, just with
different formulas for $t_{\rm m}$ and $t_{\rm e}$, and practically
constant $\t^{\rm em}(t)$ at $t<t_{\rm s}$.]

    The neglected effect of multiple reflections can be easier
analyzed in the ``pretending'' procedure with $t_{\rm s}=0$  than in
our usual procedure. Then the reflected waves carry the energy
$A_0^2\tau_{\rm bu}^{\rm rec}/2R$, and in the worst-case scenario
this wave is added in-phase with the unreceived wave with energy
$A_{\rm e}^2\tau_{\rm bu}^{\rm em}/2R$. The corresponding increase
of the energy loss is at most the factor of 2 because $\max
[|x+y|^2]/(|x|^2+|y|^2)=2$. Therefore the inefficiency $1-\eta$
cannot increase more than twice due to the neglected effect of
multiple reflections.

\vspace{0.5cm}

   Now let us estimate parameters for a possible experiment based on the
present-day technology of superconducting phase qubits. The tunable
coupler of the kind considered in Sec.\ III was used by R. Bialczak
et al. \cite{Bialczak} to create $\sigma_X^{(1)}\sigma_X^{(2)}$
coupling between two phase qubits. The coupling frequency
$\Omega_{XX}/2\pi$ was tunable between 0 and 100 MHz.
    By comparing Eq.\ (\ref{t-simple}) for $\t$ with the corresponding
two-qubit coupling frequency \cite{Bialczak,Pinto}
$\Omega_{XX}=-M/L_1L_2\omega C_{\rm qb}$ where $C_{\rm qb}$ is the
qubit capacitance,  we find $\t=2i \Omega_{XX} RC_{\rm qb}$.
 Using the values $|\Omega_{XX}|/2\pi=100$
MHz and $C_{\rm qb}= 1$ pF from the experiment \cite{Bialczak} and
$R=50\,\, \Omega$, we obtain $|\t|=0.063$. A more accurate way is to
use Eq.\ (\ref{t-lim-m}) for $\t$, that gives
$|\t|=2|\Omega_{XX}|RC_{\rm qb}/\sqrt{[1+(R/\omega
L_1)^2][1+(R/\omega L_2)^2]}$. Then for the experimental
values\cite{Bialczak} $L_1=L_2=3$ nH and $\omega /2\pi=6$ GHz, we
obtain a more accurate value $|\t|=0.053$.
  Using this value as $|\t|_{\rm max}$, assuming $\lambda /4$ resonators (so that
$\tau_{\rm rt}\approx \pi/\omega$) and $\omega / 2\pi =6$ GHz, we
find $\tau_{\rm bu}=30$ ns from Eq.\ (\ref{t_bu}); correspondingly,
the minimum duration of the procedure [see Eq.\
(\ref{duration-opt})] for $\eta =0.999$ is  $t_{\rm e}= 420$ ns.

    If we replace the two resonators with two phase qubits, then the
formalism is essentially the same, but we need to use Eq.\
(\ref{Q-qubit}) for the $Q$-factor and corresponding $\tau_{\rm
bu}$. Then expressing $\tau_{\rm bu}$ via $\Omega_{XX}$ coupling in
the experiment of Ref.\ \onlinecite{Bialczak}, we get
   $\tau_{\rm bu} =[1+(R/\omega L_1)^2]/\Omega_{XX}^2 RC_{\rm qb}$,
which gives $\tau_{\rm bu}=60$ ns for $\Omega_{XX}/2\pi =100$ MHz.
This is accidentally very close to $\tau_{\rm bu}$ for a $\lambda
/2$-resonator, because $\pi [1+(R/\omega L_{1,2})^2]/(2\omega R
C_{\rm qb})=0.996$, which is accidentally very close to 1.
Therefore, it will take a twice longer time for a transmission
between two phase qubits than in our estimate for a $\lambda
/4$-resonator: $t_{\rm e}$ for $\eta =0.999$ will be 850 ns. The
same duration is needed for a transfer between two $\lambda/2$
resonators.

    For completeness, let us write $t_{\rm e}$ in terms of $\Omega_{XX}$
explicitly, assuming identical couplers at the both sides. For a
transmission between two identical resonators the needed duration is
    \be
    t_{\rm e} = \frac{\tau_{\rm rt}[1+(\frac{R}{\omega L_1})^2]
  [1+(\frac{R}{\omega L_2})^2]}{2\Omega_{XX}^2 R^2 C_{\rm qb}^2 }
  \ln \frac{2\sqrt{\ln 4 -1}}{1-\eta} ,
    \ee
while for a transmission between two phase qubits the duration is
    \be
   t_{\rm e} =2\,\frac{1+(R/\omega L_1)^2}{\Omega_{XX}^2 RC_{\rm qb}}
    \ln \frac{2\sqrt{\ln 4 -1}}{1-\eta}.
    \ee
The formulas are different because of different relations between
the stored energy and corresponding voltage amplitude at the
coupler. The twice longer time $t_{\rm e}$ for $\lambda /2$
resonators compared with $t_{\rm e}$ for $\lambda /4$ resonators can
be understood either as because of the twice longer round-trip time
[see Eq.\ (\ref{tau-rt-simple})] or because of the twice larger
stored energy for the same amplitude of the standing wave.

 From the linear relation between $\t$ and
$\Omega_{XX}$ we see that the ON/OFF ratio discussed in this section
is the same as the ON/OFF ratio for $\Omega_{XX}$: $|\t|_{\rm
max}/|\t|_{\rm min}=|\Omega_{XX}^{\rm max}/\Omega_{XX}^{\rm min}|$.
In the experiment \cite{Bialczak} this ratio was demonstrated to be
$\sim 10^3$; therefore the ON/OFF ratio of 45 needed for $\eta
=0.999$ [see Eq.\ (\ref{ON/OFF-sym})] is easily realizable
experimentally.

In the derivation we assumed exactly equal frequencies of the two
resonators. Let us very crudely estimate the effect of a small
detuning $\delta \omega$. Assuming exact cancellation of the
reflected wave at time $t_{\rm m}$ and $\tau_{\rm bu}^{\rm
em}=\tau_{\rm bu}^{\rm rec}$, we estimate the amplitude of the
back-reflected wave at $t\neq t_{\rm m}$ as $A_{\rm m} e^{-|t-t_{\rm
m}|/2\tau_{\rm bu}} \delta \omega |t-t_{\rm m}|$, so that the
corresponding energy loss is $\int_{-\infty}^{\infty} A_{\rm m}^2
(\delta\omega)^2 t^2 e^{-|t|/\tau_{\rm bu}} dt/2R= 4 A_{\rm m}^2
(\delta \omega)^2 \tau_{\rm bu}^3/2R$. Comparing it with the total
transmitted energy $2A_{\rm m}^2\tau_{\rm bu}/2R$ we obtain the
relative loss $1-\eta_{\delta\omega}=2 (\delta\omega \, \tau_{\rm
bu})^2$. Therefore for the required total efficiency $\eta$ we can
tolerate detuning of about
    \be
|\delta \omega|/2\pi \alt \frac{\sqrt{1-\eta}}{10\, \tau_{\rm bu}}.
    \label{delta-omega}\ee
 For the above example with $\lambda/4$ resonators and $\eta =0.999$
this gives a tolerable detuning of only 0.1 MHz. Such a strict
requirement on the detuning means that in an experiment at least one
resonator should be slightly tunable in frequency. Moreover, as
follows from Eqs. (\ref{t-gen-m})--(\ref{r_l-gen-m}), modulation of
the transparency $\t$ leads to a slight change of the reflection
phase and therefore a slight change of the resonator frequency,
which is nevertheless sufficient to violate the requirement
(\ref{delta-omega}). A compensation for this effect by a slight
tuning of the resonator frequency is necessary for a high-fidelity
transfer, at least for the physical scheme considered in Sec.\ III.

    Since our formalism assumes high-$Q$ resonators, let us estimate
the limitation on the maximum transparency $|\t_{\rm max}|$ needed
for validity of our theory to achieve an efficiency $\eta$. In the
theory we assume an instantaneous change of a resonator amplitude,
while physically it takes time of about $\tau_{\rm rt}$. This leads
to a relative error of about $|\t|^2$ for the amplitude of the
emitted wave. Therefore the reflection cancellation may be imperfect
up to an amplitude $|A(t)|(|\t^{\rm em}(t)|^2+|\t^{\rm rec}(t)|^2)$,
and the corresponding energy dissipation can be estimated as
$|A_{\rm m}|^2 |\t|_{\rm max}^4\tau_{\rm bu}/2R$, leading to
additional inefficiency $1-\eta_Q \sim |\t|_{\max}^4$. This means
that our high-$Q$ theory becomes invalid when $|\t|_{\rm max}^4 \agt
1-\eta$. A similar estimate can be obtained when taking into account
multiple reflections of amplitude $\sim A_0$. Then extra energy loss
is $\sim |A_0 A_{\rm m}|\, |\t|_{\rm max}^2 \tau_{\rm bu}/2R$,
leading to $1-\eta_Q \sim |\t|_{\rm max}^2\sqrt{1-\eta}$, which also
makes the theory invalid when $|\t|_{\rm max}^4 \agt 1-\eta$.
Therefore, for the couplers with $|\t|_{\rm max}<0.1$ our theory
should remain valid up to efficiencies of about $\eta \simeq
0.9999$.

\section{Conclusion}

In this paper we have proposed and analyzed a method to transmit
classical microwave energy as well as a logical qubit between two
same-frequency resonators with an efficiency $\eta$ arbitrarily
close to 100\%. This is done using two tunable couplers at both ends
of the transmission line (Fig.\ 1), which modulate the coupling in a
specific way (Fig.\ 2). In the first half of the procedure the
receiving coupling is kept maximum while the emitting coupling
increases in time, and in the second half the emitting coupling is
kept maximum, while the receiving coupling decreases. The main idea
is to cancel the wave reflected back into the transmission line from
the receiving end at any time except for the initial period $t_{\rm
s}$.

    The required ON/OFF ratios for the couplers and duration
$t_{\rm e}$ of the procedure excluding the propagation time depend
on the desired efficiency $\eta$. The main results for these
requirements in the symmetric case are given by Eqs.\
(\ref{duration-opt}) and (\ref{ON/OFF-sym}). Compared with the
proposal of Ref.\ \onlinecite{Yurke} in which only the emitting
coupler is modulated, our procedure requires much shorter duration
[crudely by the factor $1/(1-\eta)$] and much smaller ON/OFF ratio
[crudely by the factor $1/\sqrt{1-\eta}$]. This hopefully makes
realistic an experiment on flying qubits using the superconducting
phase qubit technology.

    Our results show that a transfer with $\eta =0.999$
(excluding losses due to dissipation) can be performed using the
tunable coupler of Ref.\ \onlinecite{Bialczak} with parameters
corresponding to varying the two-qubit coupling frequency between
$\Omega_{XX}^{\rm max}/2\pi \simeq 100$ MHz and $\Omega_{XX}^{\rm
min}/2\pi \simeq 2.2$ MHz (this was already
demonstrated\cite{Bialczak}), and the transfer procedure requires
duration $t_{\rm e}\simeq 420$ ns if $\lambda /4$ resonators are
used. This experiment would tolerate only a very small detuning
between the resonator frequencies; a crude estimate is $\delta
\omega/2\pi \alt$ 0.1 MHz for $\omega/2\pi =6$ GHz and $\eta
=0.999$. The transfer can also be made between two phase qubits
directly connected to a transmission line; however this would
require approximately twice longer duration (almost the same as for
$\lambda /2$ resonators) and would significantly suffer from the
qubit decoherence. The results are practically the same for a
somewhat modified (more symmetrical) procedure described by Eqs.\
(\ref{t-em})--(\ref{tmte}). While in this paper we focused on the
superconducting phase qubit technology, the general procedure of the
quantum state transfer is applicable to other realizations
(including optical photons), in which tunable couplers can be
realized. For such realizations the results of Sec.\ III for the
transmission and reflection amplitudes of the coupler should be
replaced by the corresponding results for a different tunable
coupler, while other results remain unchanged.

    We emphasize that our analysis was essentially classical,
but following the formalism of Ref.\ \onlinecite{Yurke} we do not
expect that the results of a fully quantum analysis could be
significantly different. Nevertheless, such work would definitely be
useful in future, especially taking into account a weak nonlinearity
of resonators due to coupling with qubits. There are also other
interesting questions for further study. In particular, it is
important to study numerically the effects of the procedure
imperfections, including weak detuning, imperfect timing, and
deviations from the ideal time dependence of the coupling
modulation. Also, we have analyzed only the weak-coupling case and
used an assumption of fixed frequency, while it would be interesting
to do a numerical analysis for tunable couplers with moderate
coupling using Eqs.\ (\ref{t-gen-m}) and (\ref{r_l-gen-m}), which
take into account the phase modulation of the $S$-parameters.
Another important issue is the effect of multiple reflections in a
short transmission line.
 Explicit account for the
energy dissipation in the resonators and transmission line would
also be relevant to an experiment on flying microwave qubits, which
can hopefully be realized soon.

\vspace{0.3cm}

    The author thanks Andrei Galiautdinov, Michael Geller, Andrew
Cleland, and John Martinis for attracting interest to this problem
and useful discussions. The author also thanks Kyle Keane and John
Martinis for critical reading of the manuscript. The work was
supported by NSA/IARPA/ARO Grant No. W911NF-10-1-0334.


\begin{thebibliography}{99}
\bibitem{note-evol} Equation (\ref{B-evol}) has a transparent
physical meaning. The $A$-term has the phase factor $e^{i\varphi}$
acquired from the round trip (excluding reflection at the barrier)
because the $B$-amplitude is defined as incident to the barrier (not
outgoing, see Fig.\ 1). The total round-trip factor is then $\r_2
e^{i\varphi}$, which enters the $B$-term. The time constant
$\tau_{\rm rt}$ has a clear meaning of the round-trip time in the
ideal case of a long resonator, while in the case of a significant
energy storage at the boundaries it is better to define $\tau_{\rm
rt}$ via the energy-amplitude relation like in Eq.\
(\ref{tau-rt-def}). Note that in Eq.\ (\ref{B-evol}) we assume the
limit of high $Q$ and small detuning ($|\r_2 e^{i\varphi}-1|\ll 1$),
and in this limit Eq.\ (\ref{B-evol}) is actually not the unique
description of the evolution.
\end{thebibliography}
\end{document}